\documentclass[prc,aps,showpacs]{revtex4}
\usepackage{graphicx}
\begin{document}

\title{Effect of $\Sigma$-beam Asymmetry Data on Fits to
	Single Pion Photoproduction off Neutron}

\author{I.I.~Strakovsky\footnote[1]{Electronic address: igor@gwu.edu},
	R.A.~Arndt\footnote[2]{Deceased},
	W.J.~Briscoe,
	M.W.~Paris,
	R.L.~Workman}
\affiliation{The George Washington University, Washington, DC 20052, USA}

\begin{abstract}
We investigate the influence of new GRAAL $\Sigma$-beam asymmetry 
measurements on the neutron in multipole fits to the single-pion 
photoproduction database. Results are compared to those found with 
the addition of a double-polarization quantity associated with the
sum rule.
\end{abstract}

\pacs{13.60.Le, 25.20.Lj, 13.88.+e, 11.80.Et}

\maketitle

\section{Introduction}

Only with good data on both proton and neutron targets one can hope 
to disentangle the isoscalar and isovector EM couplings of the 
various baryon resonances, as well as the isospin properties of the 
non-resonant background amplitudes.  In particular, the simple quark 
model predicts several resonances that couple much stronger to the 
neutron than to the proton.  The lack of $\gamma n\to\pi^-p$ and 
$\gamma n\to\pi^0n$ data does not allow us to be as confident about 
the determination of neutron couplings relative to those of the 
proton.

Some of the N$^\ast$ baryons, N(1675)5/2$^-$, for instance, have 
stronger electromagnetic couplings to the neutron than to the proton 
but parameters are very uncertain. PDG~\cite{PDG} estimate for the 
A$_{1/2}$ and A$_{3/2}$ decay amplitudes of the N(1720)3/2$^+$ state 
are consistent with zero, while the recent SAID 
determination~\cite{Dugger} gives small but non-vanishing values.  
The reason for the disagreement between the PDG estimate for the 
A$_{1/2}$ decay amplitude and the recent SAID 
determination~\cite{Dugger} is also unclear. Other unresolved issues 
relate to the N(1700)$3/2^-$ and second P$_{11}$, N(1710)1/2$^+$, 
that are not seen in the recent GW $\pi$N partial-wave analysis 
(PWA)~\cite{Ar06}, contrary to other PWAs used by the Particle Data 
Group~\cite{PDG}.

New, high quality data on $\gamma n\to\pi^-p$ and $\gamma n\to\pi^0n$ 
are needed to shed light on these issues, and the tagged-photon hall 
at GRAAL offered a state-of-the-art facility to obtain such data.  
Here we report on an analysis included novel $\Sigma$-measurements, 
covering incident photon energies from threshold ($E_\gamma = 707$~MeV) 
up to $E_\gamma = 1500$~MeV.  The present measurement of $\Sigma$s 
for $\vec{\gamma}n\to\pi^-p$~\cite{Ma10} and for 
$\vec{\gamma}n\to\pi^0n$~\cite{Sa09} is part of an extensive program 
at the GRAAL to provide data of unrivaled quality on charged and 
neutral meson photoproduction on the neutron, which includes 
polarized beam observable in addition to the cross sections.

\section{New GRAAL Measurements for $\Sigma$ on the Neutron}

\begin{figure}
\centerline{
\includegraphics[height=0.95\textwidth, angle=90]{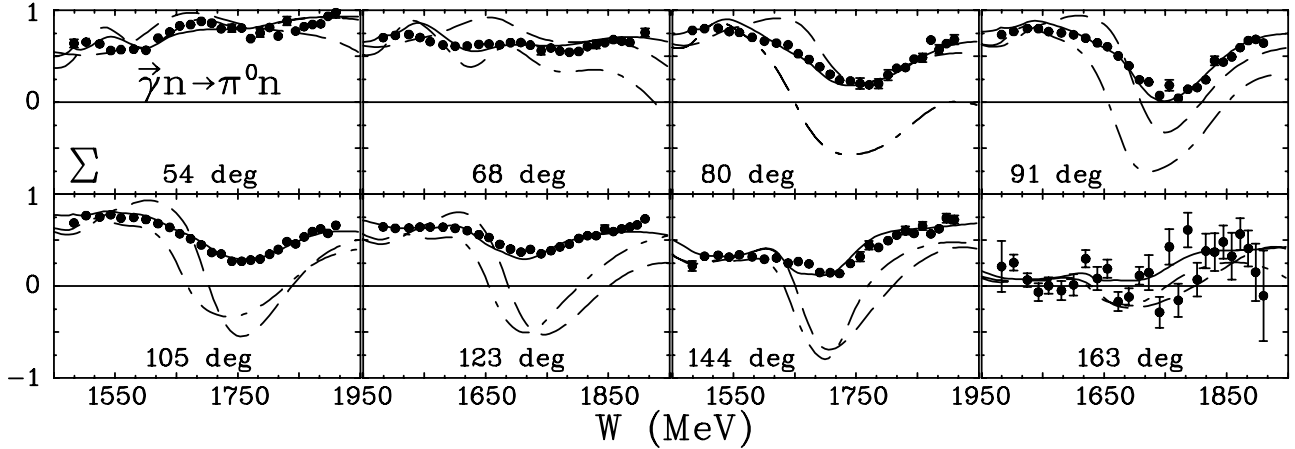}}
\vspace{3mm} \caption{$\Sigma$-beam asymmetry for
        $\vec{\gamma}n\to\pi^0n$. Data from GRAAL
        Collaboration~\protect\cite{Sa09}. Solid lines 
	correspond to the SAID-MA09 solution (GRAAL data 
	included in the database)~\protect\cite{Ma10}. 
	Dash-dotted (dashed) lines show the 
	SAID-SP09~\protect\cite{Dugger}
        (MAID2007~\protect\cite{maid}) (GRAAL data excluded 
	in the database). \label{fig:g1}}
\end{figure}

To gauge the influence of new GRAAL data and their compatibility 
with previous measurements, the GRAAL $\Sigma$s have been 
included in a number of fits using the full SAID database
for $\gamma N\to\pi N$ up to $E_\gamma=2.7$~GeV~\cite{said}.  
The impact of new data on the SAID PWA can be understood 
from the comparison of the new SAID fit MA09~\cite{Ma10}, 
which involves new GRAAL data, with the previous SAID fit 
SP09~\cite{Dugger} and MAID2007 results~\cite{maid}.

216 $\Sigma$s for $\pi^0n$ final state at E$_\gamma$=703--
1475~MeV and $\theta$=53--164$^\circ$ with 99 $\Sigma$s for 
$\pi^-p$ final state at E$_\gamma$=753--1439~MeV and 
$\theta$=33--163$^\circ$ GRALL data have been added to the 
GW SAID database~\cite{said}.  We have to notice that this 
GRAAL $\pi^0n$ contribution doubled the World database for 
this reaction.  Our best fit MA09~\cite{Ma10} for $\pi^0n$ 
and $\pi^-p$, reduced initial $\chi^2$/dp=223 and 
89 (SP09~\cite{Dugger}) to 3.1 and 4.9, respectively.  It 
shows, in particular, that previous $\pi^-p$ measurements 
provided a better constraint vs. $\pi^0n$ case.

\begin{figure}
\centerline{
\includegraphics[height=0.95\textwidth, angle=90]{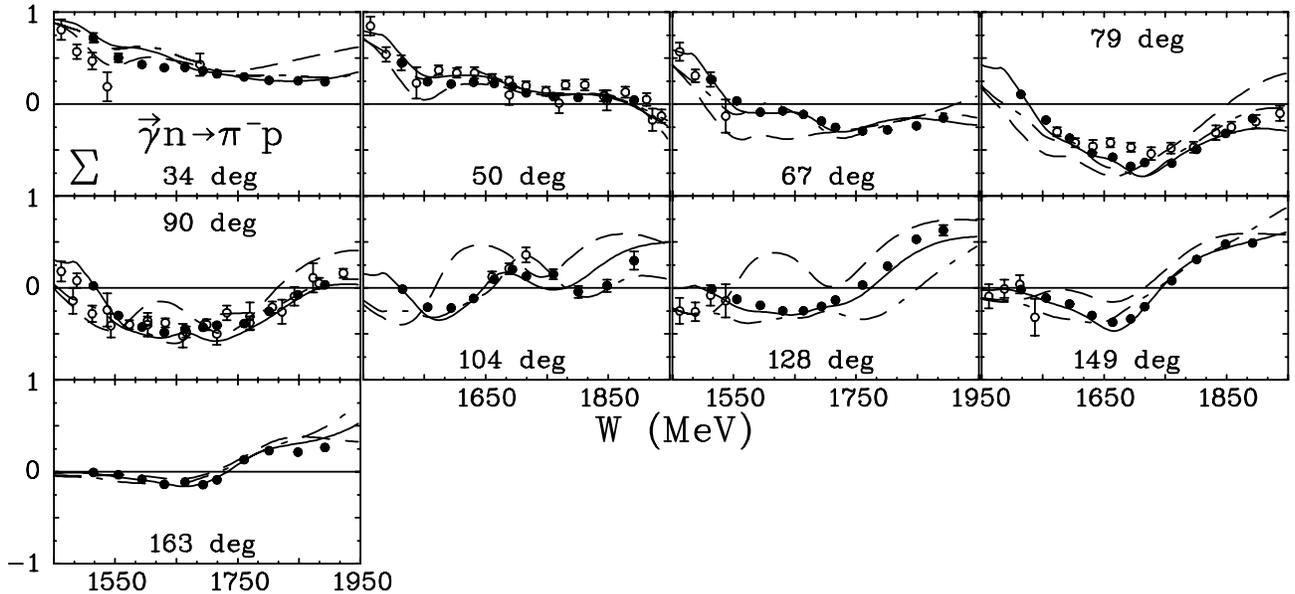}}
\vspace{3mm} \caption{$\Sigma$-beam asymmetry for 
	$\vec{\gamma}n\to\pi^-p$. Data from GRAAL 
	Collaboration~\protect\cite{Ma10}. The notation of the 
	PWA solutions is the same as in Fig.~\protect\ref{fig:g1} 
	\label{fig:g2}}
\end{figure}

In Figs.~\ref{fig:g1} and \ref{fig:g2}, we show the 
excitation functions for several production angles.  The 
number of the distributions shown is enough to illustrate 
the quality of new GRAAL data, the main features of the 
$\gamma n\to\pi N$ dynamics at the measured energy range, 
and the impact of the present data on PWAs.  The most 
noticeable effect of the present data on the new MA09 is 
due to very good measurements of the medium-angle 
(65--140$^\circ$) $\Sigma$s for W in the range above 
1650~MeV. Earlier, this angular region either had been 
measured with worse accuracy or could only be reached by 
extrapolation.

\begin{figure}
\centerline{
\includegraphics[height=0.4\textwidth, angle=90]{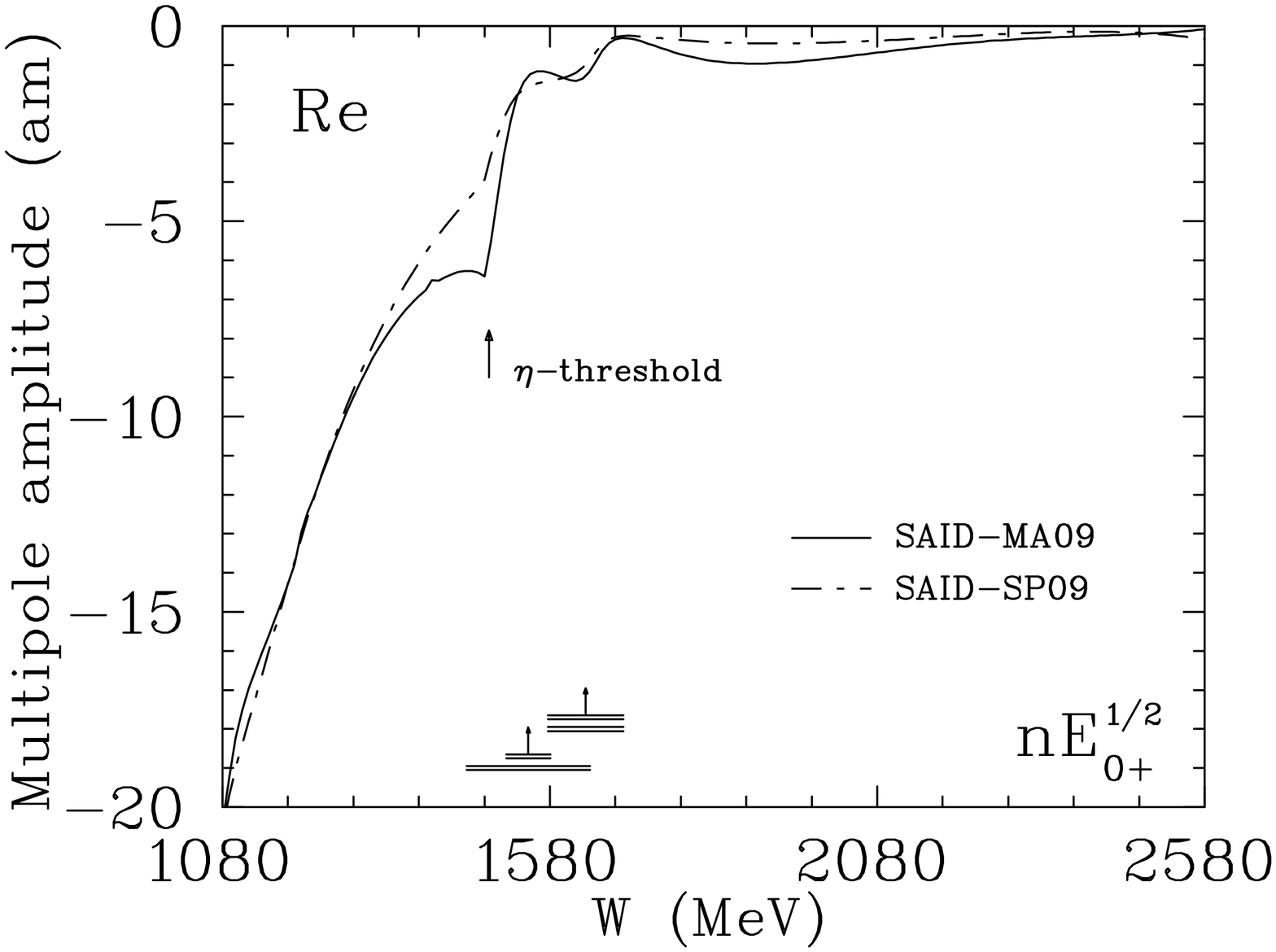}\hfill
\includegraphics[height=0.4\textwidth, angle=90]{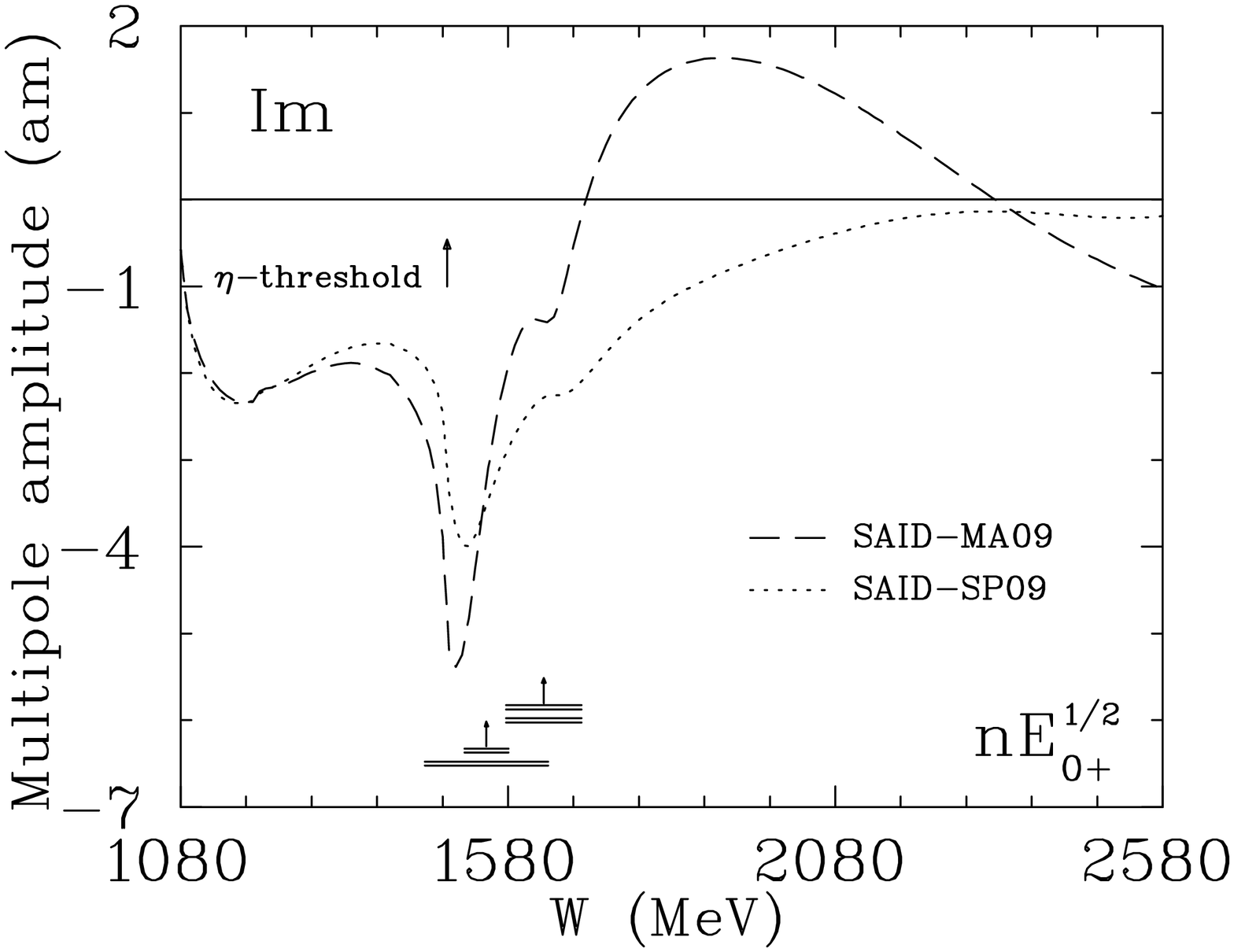}}
\vspace{3mm} \caption{The multipole amplitude $S_{11}$nE
        ($_nE_{0+}^{1/2}$). (a) Re and (b) Im parts show.
	The vertical arrows indicate $W_R$ (Breit-Wigner 
	mass) and the horizontal bars show the full and 
	partial width $\Gamma$ and $\Gamma_{\pi N}$
	associated with the SAID solution SP06 for
	$\pi N$~\protect\cite{Ar06}. \label{fig:g3}}
\end{figure}

The difference between our MA09 and SP09 results for the neutron 
target is visible specifically for $S_{11}$nE (Fig.~\ref{fig:g3}).
It is observed above E$_\gamma\sim$400~MeV while modified MAID2007 
shown a significant changes vs. MAID2007~\cite{Sa09} above 1~GeV 
(see Fig.~7 at Ref.~\cite{Sa09}).

The difference between previous pion photoproduction and new GRAAL 
measurements may result in significant changes in the neutron 
couplings.

\section{Helicity-Dependent Photoabsorption Cross Sections on 
	the Neutron}

The amplitudes obtained in our analyses can be used to evaluate the 
single-pion production component of several sum rules, in particular
GDH, Baldin, and forward spin polarizability~\cite{Ar02}.  
In Table~\ref{tab:a}, we summarized our results for the neutron 
target.

\begin{figure}
\centerline{
\includegraphics[height=0.35\textwidth, angle=90]{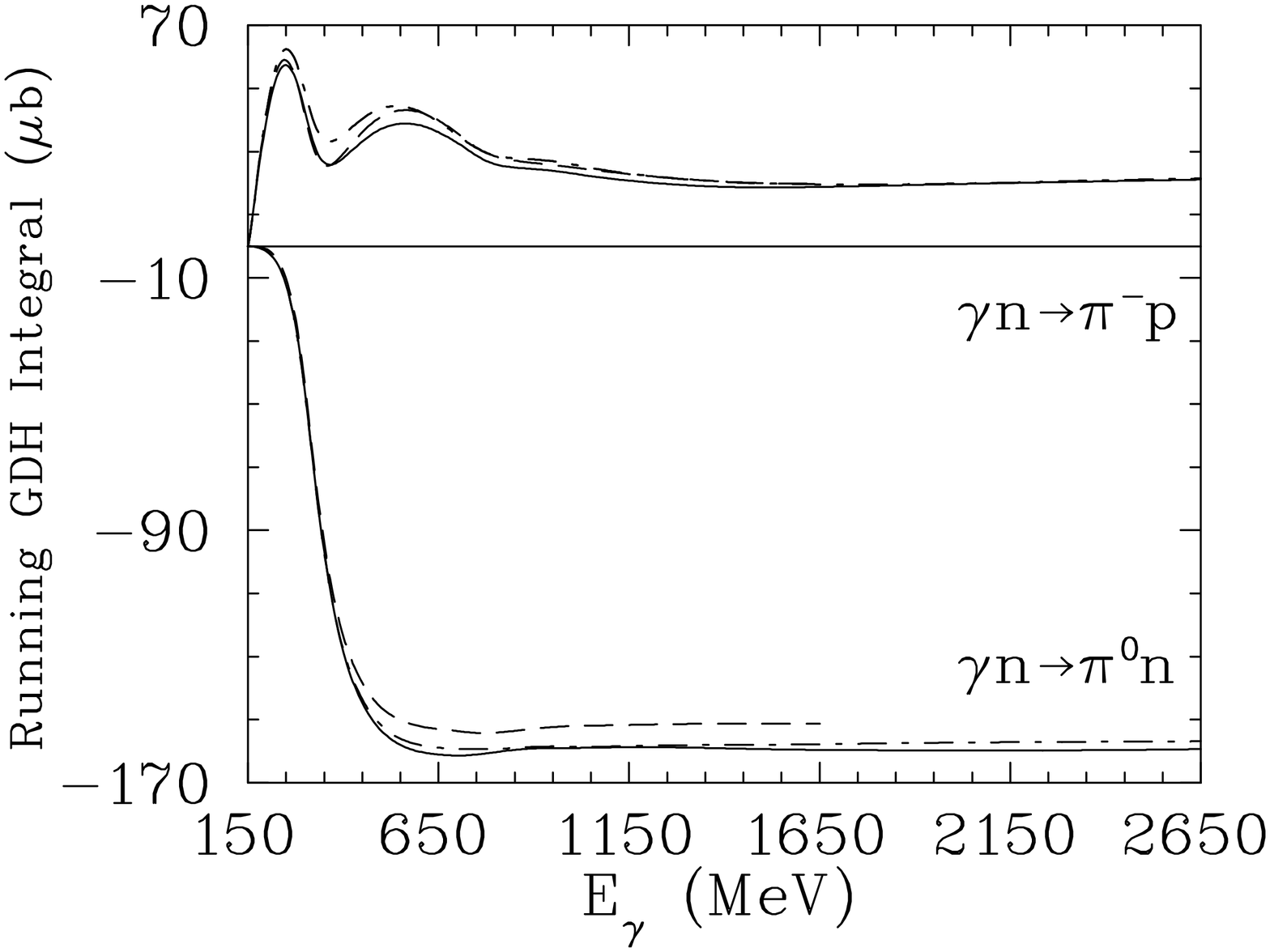}\hfill
\includegraphics[height=0.31\textwidth, angle=90]{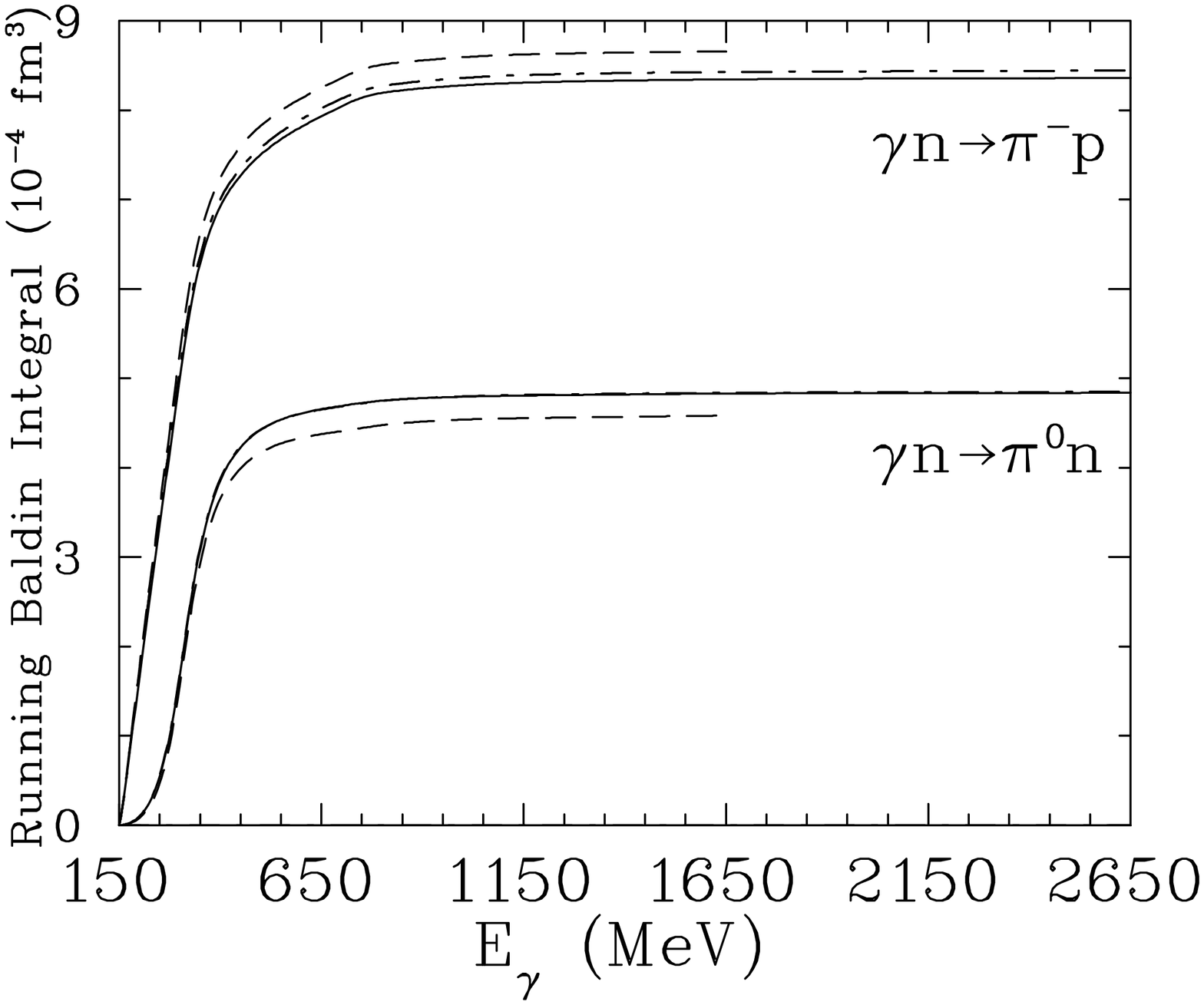}\hfill
\includegraphics[height=0.34\textwidth, angle=90]{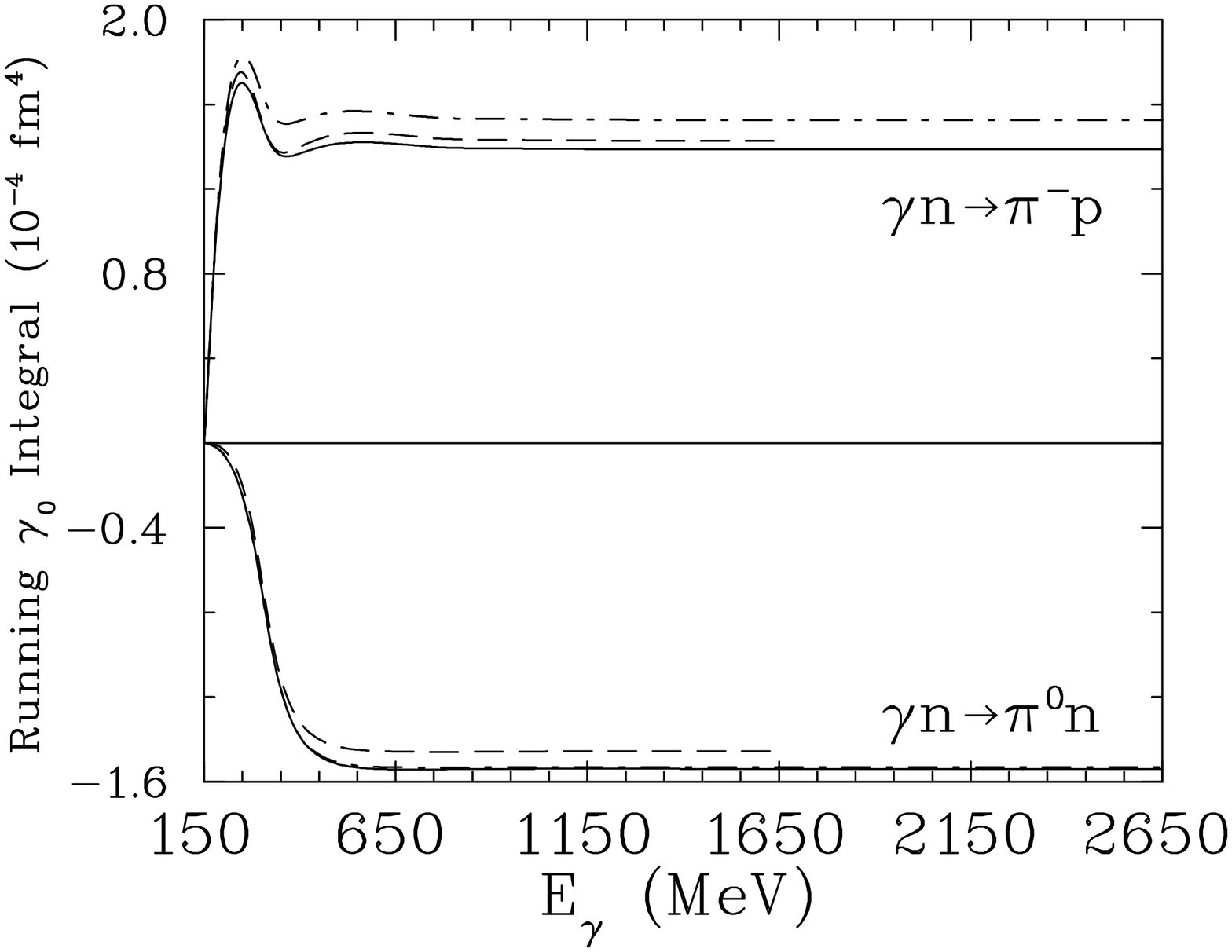}}
\vspace{3mm} \caption{Running (a) GDH, (b) Baldin, and (c) forward
        spin polarizability $\gamma_0$ integrals for the neutron
        target.  The solid (dash-dotted) lines represent the
        SAID-MA09 (SAID-SP09) solution.  Dashed lines show the
        MAID2007 predictions. \label{fig:g4}}
\end{figure}

The running integrals are shown in Fig.~\ref{fig:g4}.  The 
evaluation of sum rules (GDH, Baldin, and forward spin polarizability) 
for the neutron target and for a single pion contribution exhibits 
convergence by 1~GeV. Agreement with Mainz is good.  Clearly, 
calculations above 450~MeV have to take into account contributions 
beyond single-pion photoproduction.

\begin{table}
\caption{Comparison of the recent SAID-MA09, SAID-SP09,
         and MAID2007 calculations for the GDH, Baldin
         and the forward spin polarizability from threshold
         up to 2.5~GeV in W (for MAID up to 2~GeV) and
         displayed as MA09/SP09/MAID2007.} \label{tab:a}
\vspace{2mm}
\begin{tabular}{|c|c|c|c|}
\colrule
Reaction  &    GDH    &   Baldin       & $\gamma_0$ \\
  & ($\mu b$) &($10^{-4} fm^3$)& ($10^{-4} fm^4$) \\
\colrule
$\gamma n\to\pi^-p$ & 21/ 21/ 20           & 8.4/8.4/8.7 & 1.4/ 1.5/ 1.4 \\
$\gamma n\to\pi^0n$ & $-$159/$-$157/$-$151 & 4.8/4.8/4.6 & $-$1.5/$-$1.5/$-$1.5 \\
\colrule
\end{tabular}
\end{table}

\acknowledgments
This work was supported in part by the U.~S.~Department of Energy 
under Grant DE-FG02-99ER41110.


\end{document}